# Error Field and Correction Coils in DTT: a preliminary analysis

Raffaele Albanese[1,2,3], Tommaso Bolzonella[4], Andrea G. Chiariello[2,5], Antonio Cucchiaro[3], Antonio Iaiunese[1], Alessandro Lampasi[3,6], Raffaele Martone[2,3], Lidia Piron[4,7], Aldo Pizzuto[3], and Pasquale Zumbolo[1]

[1] *Università degli studi di Napoli Federico II, via Claudio 21, I-80124 Napoli, Italy*
[2] *Consorzio CREATE c/o Università degli studi di Napoli Federico II, via Claudio 21, I-80124 Napoli, Italy*
[3] *DTT S.C. a r.l. Via E. Fermi 45 I-00044, Frascati (Rome), Italy*
[4] *Consorzio RFX, Corso Stati Uniti 4, 35127, Padova, Italy*
[5] *Università degli studi della Campania, via Roma, 29, I-81031 Aversa (CE), Italy*
[6] *ENEA, Frascati (Rome), Italy*
[7] *Università degli Studi di Padova, Padova, Italy*

*Abstract*— The Divertor Tokamak Test (DTT) facility, construction starting at Frascati, Italy, is designed to test different solutions for divertor in view of DEMO. A preliminary analysis of the error fields (EFs) assumed a simplified model of rigid and independent displacements and rotations. A methodology based on the first order truncated Taylor expansion has been applied, linking the displacement parameters and the EFs within the required accuracies. A system of in-vessel copper coils has been designed to counteract EFs and the ampere-turns necessary to force them back within the request limits has been calculated. Here, the details of the analysis have been provided.

## 1. Introduction

A magnetic field with prescribed specifications is necessary for the plasma confinement in a Tokamak fusion device. Deviations from nominal magnetic field map, due to external perturbations, are known in literature as Error Fields (EFs) [1][2]. Inaccuracies and tolerances in manufacturing and assembly of magnets, joints, current feeds, or non-axially symmetric parts, are sources of EFs. To prevent consequences on the plasma stability, the EFs can be reduced by means of Error Field Correction Coils (EFCCs).

Here, a preliminary analysis, considering simplified models for the manufacturing and assembly inaccuracies, such as independent rigid displacement, rotations, and deformations of coils, has been developed and validated for the DTT (Divertor Tokamak Test) device [3] to study the impact of the EFs and design an EFCCs system able to reduce the perturbations below a specific threshold. The characteristic dimensions of the DTT magnets are in the order of meters, with tolerances in the order of millimeters. Thus, a model, based on first order Taylor approximation, is accurate enough to be used to produce a statistical analysis, evaluate the impact on the EFs and optimize the EFCCs currents to counteract them.

## 2. Mathematical Model

### 2.1 EFs quantitative description

In Tokamak nominal operations, the flux density cartesian component $B_{x0}(\boldsymbol{r})$ (similarly for $y$ and $z$ components) is the superposition of $N$ nominal sources contributions, including magnetic system and plasma:

$$B_{x0}(\boldsymbol{r}) = B_{x0}^{(1)}(\boldsymbol{r}) + B_{x0}^{(2)}(\boldsymbol{r}) + \ldots + B_{x0}^{(N)}(\boldsymbol{r}) \quad (1)$$

where $\boldsymbol{r}$ is the generic field point in $\mathbb{R}^3$ and $N$ the number of sources.

The EFs act as an additive contribution $\Delta B_x(\boldsymbol{r})$, altering the field in (1), giving the perturbed field $B_x(\boldsymbol{r})$:

$$B_x(\boldsymbol{r}) = B_{x0}(\boldsymbol{r}) + \Delta B_x(\boldsymbol{r}) \quad (2)$$

where $\Delta B_x(\boldsymbol{r})$ takes account of $N$ perturbative contributions corresponding to the deformation of the $N$ sources:

$$\Delta B_x(\boldsymbol{r}) = \Delta B_x^{(1)}(\boldsymbol{r}) + \Delta B_x^{(2)}(\boldsymbol{r}) + \ldots + \Delta B_x^{(N)}(\boldsymbol{r}) \quad (3)$$

The amplitude of the $i$-th source perturbations can be related to a suitable set of a finite dimensional space parameters $\underline{\Delta p}^{(i)} = \left[\Delta p_1^{(i)}, \Delta p_2^{(i)}, \ldots \Delta p_K^{(i)}\right]$ with $i$=1, ..., $N$, and $K$ the dimension of the space. It should be noted that $\Delta B_x^{(i)}(\boldsymbol{r}, \underline{\Delta p}^{(i)})$ may depend non-linearly on $\underline{\Delta p}^{(i)}$.

A standard figure of merit known in literature as TMEI (Three Mode Error Index) [4] can be used to take into account EFs impact:

$$TMEI = \frac{\sqrt{0.2\|B_{1,1}\|_2 + \|B_{1,2}\|_2 + 0.8\|B_{1,3}\|_2}}{B_{tor}} \quad (4)$$

where $B_{tor}$ is the unperturbed toroidal field on the magnetic axis of the device [5][6], while $B_{1,1}$, $B_{1,2}$ and $B_{1,3}$ are the spectral components of *3D* complex Fourier space [7] defined as:

$$B_{n,m} = \frac{1}{(2\pi)^2} \oiint B_N(\theta, \varphi) e^{-i(n\varphi - m\tilde{\theta})} d\tilde{\theta} \, d\varphi \quad (5)$$

with $m$=1,2,3, $n = 1$, $B_N(\theta, \varphi)$ the normal component of the field to the q=2 surface, $\theta, \varphi$ the angular coordinates of a quasi-toroidal ($\rho, \theta, \varphi$) coordinate system, and $\tilde{\theta}$ the angular curvilinear abscissa [8], defined as:

$$\tilde{\theta}(l) = \frac{1}{q} \int_0^l \frac{B_t}{r B_p} dl \quad (6)$$

where $l$ is the length of the field line projected on the poloidal section and $B_t$ and $B_p$ are the toroidal and poloidal components of the field on the q=2 surface.

### 2.2 Linearized approach

The $i$-th term in (3) can be expressed as a Taylor series respect to the perturbation parameters, truncated at first order derivatives:

---

*Andreagaetano.Chiariello@unicampania.it*

$$\Delta B_x^{(i)} \simeq \underline{\Delta p}^{(i)} \bullet \nabla B_x^{(i)} \quad (7)$$

Each source can be discretized in a finite number of current segments $N_{seg}$ [9][10]. The field of the discretized source is the superposition of its segments' fields contributions, and, generally, the higher the discretization, the higher the accuracy.

Since the analytical expression of the field of each segment depends on the coordinates of its extremes and on its current, the derivatives in (7) can be split by using the chain rule.

In a finite dimensional space, the expression in (7), including $y$ and $z$ fields components, becomes a matrix relation between $\Delta \boldsymbol{B}^{(i)}(\boldsymbol{r}) \in \mathbb{R}^3$ and the vector $\underline{\Delta p}^{(i)} \in \mathbb{R}^K$ through a matrix $\underline{\underline{M}}^{(i)} = \underline{\underline{M}}_S^{(i)} \bullet \underline{\underline{M}}_G^{(i)}$, where $\underline{\underline{M}}_S^{(i)} \in \mathbb{R}^{3 \times 3N_{seg}}$ and $\underline{\underline{M}}_G^{(i)} \in \mathbb{R}^{3N_{seg} \times K}$ are related to chain rule. Using (5), the contribution to the harmonics due to the i-th source becomes:

$$\underline{B}_{harms}^{(i)} = \begin{bmatrix} B_{1,1}^{(i)} \\ B_{1,2}^{(i)} \\ B_{1,3}^{(i)} \end{bmatrix} = \underline{\underline{\widetilde{M}}}^{(i)} \begin{bmatrix} \Delta p_1^{(i)} \\ \Delta p_2^{(i)} \\ \vdots \\ \Delta p_K^{(i)} \end{bmatrix} \quad (8)$$

where $\underline{\underline{\widetilde{M}}}^{(i)} \in \mathbb{C}^{3 \times K}$ is a i-th complex matrix linking the Fourier harmonics to the perturbed parameters space. The harmonics vector $\underline{\widetilde{B}}_{harms}$, due to all the sources, is the superposition of all contributions in (8):

$$\underline{\widetilde{B}}_{harms} = \underline{\underline{\widetilde{M}}}^{(1)} \underline{\Delta p}^{(1)} + \underline{\underline{\widetilde{M}}}^{(2)} \underline{\Delta p}^{(2)} + \ldots + \underline{\underline{\widetilde{M}}}^{(N)} \underline{\Delta p}^{(N)} \quad (9)$$

The effectiveness of the method has been successfully verified by comparing the "linearized field" to the one due to the direct numerical evaluation, obtaining the required accuracy (about $1\mu T$).

Once the set of perturbation parameters is defined, the TMEI modes matrices $\underline{\underline{\widetilde{M}}}^{(1)}, \ldots, \underline{\underline{\widetilde{M}}}^{(N)}$ in (9) can be stored in advance; this allows a very fast stochastic analysis in the perturbed parameters space (only matrix by vector multiplications needed).

*2.3 EFCCs current optimizations*

Once the number of cases $N_{cases}$ has been chosen, a distribution of harmonics' values $\underline{\widetilde{B}}^c_{harms}$, with $c = 1, \ldots N_{cases}$, can be obtained from (9), by randomly varying the vector $\underline{\widetilde{\Delta p}}^c = [\Delta p^{(1)}, \Delta p^{(2)}, \ldots, \Delta p^{(N)}]$ in the range of each perturbation amplitude and for each source.

Each examined case $\underline{\widetilde{B}}^c_{harms}$ can be used as the input of a EFCCs currents optimization tool solving the system:

$$\underline{\underline{\widetilde{G}}} \, \underline{I}_{CC}^c = \underline{\widetilde{B}}^c_{harms} \quad (10)$$

where $\underline{\underline{\widetilde{G}}}$ is a $\mathbb{C}^{3 \times n_{EFCCs}}$ matrix linking the harmonics for unitary currents and the $n_{EFCCs}$ feeders $\underline{I}_{CC}^c$. Since, typically $n_{EFCCs} > 3$, the routine acts on the underdetermined system and finds a solution minimizes the maximum EFCCs currents used to reduce the overall EF below the required threshold. System (10) is then replaced by the minimization of a quadratic function:

$$\min_{I_{CC}} \frac{1}{2} (\underline{I}_{CC}^c)^T \underline{\underline{\widetilde{G}}}^T \underline{\underline{W}}^T \underline{\underline{W}} \, \underline{\underline{\widetilde{G}}} \, \underline{I}_{CC}^c - \left( (\underline{\widetilde{B}}^c_{harms})^T \underline{\underline{W}} \underline{\underline{W}} \, \underline{\underline{\widetilde{G}}} \right)^T \underline{I}_{CC}^c \quad (11)$$

with $lb \le \underline{I}_{CC}^c \le ub$

where $\underline{\underline{W}}$ is a suitable regularization matrix and $lb$ and $ub$ are vectors with the current limits. The MATLAB tool *quadprog* [11], has been used to solve (11). Finally, the percentage of corrected cases can be obtained.

## 3. DTT: Description of coil perturbations

DTT is the Italian high field (6 T), high current (5.5 MA) super-conducting Tokamak [3], designed to explore new solutions for the power and particle exhaust.

*3.1 DTT magnetic and EFCCs systems description*

The DTT PF/CS system design is still ongoing. The set of geometrical parameters used is illustrated in Table I. Since they are symmetric with respect to the equatorial plane, only the upper side coils have been reported.

Table I. Main PF/CS system parameters

| Name | $R_B$ [m] | $Z_B$ [m] | DR [m] | DZ [m] | nR | nZ |
|---|---|---|---|---|---|---|
| PF1 | 1.4000 | 2.7600 | 0.5100 | 0.5904 | 18 | 20 |
| PF2 | 3.0795 | 2.5340 | 0.2790 | 0.5168 | 10 | 16 |
| PF3 | 4.3511 | 1.0150 | 0.3898 | 0.4522 | 14 | 14 |
| CS3U-H | 0.4896 | 2.1658 | 0.1213 | 0.7880 | 4 | 17 |
| CS3U-M | 0.5960 | 2.1658 | 0.0915 | 0.7880 | 4 | 20 |
| CS3U-L | 0.6935 | 2.1658 | 0.1035 | 0.7880 | 6 | 24 |
| CS2U-H | 0.4896 | 1.2994 | 0.1213 | 0.7880 | 4 | 17 |
| CS2U-M | 0.5960 | 1.2994 | 0.0915 | 0.7880 | 4 | 20 |
| CS2U-L | 0.6935 | 1.2994 | 0.1035 | 0.7880 | 6 | 24 |
| CS1U-H | 0.4896 | 0.4331 | 0.1213 | 0.7880 | 4 | 17 |
| CS1U-M | 0.5960 | 0.4331 | 0.0915 | 0.7880 | 4 | 20 |
| CS1U-L | 0.6935 | 0.4331 | 0.1035 | 0.7880 | 6 | 24 |

In Tab. I, $R_B$ and $Z_B$, DR and DZ, nR and nZ are the barycentre coordinates, the sizes of the sections, and the number of turns along R and Z, respectively of each axisymmetric coil.

The TF system, with 18 identical D-shaped coils, produces a toroidal field of 6 T on the device axis [12].

An EFCC system, consisting in three arrays ("lower", "equatorial" and "upper") with 9 identical independent filamentary copper coils each, is arranged on the plasma side of the vessel (Table II) to correct the three poloidal modes of TMEI. The model of the 9 equatorial EFCCs is up-down symmetric even if in the final design a slight asymmetry does appear.

The poloidal cross section of PF/CS and TF coils system is reported in Fig. 1.a while a 3D representation of the EFCC system in Fig. 1.b. The upper $R_U$, $Z_U$ and lower $R_L$, $Z_L$ coordinates of each array are reported in Tab II.

Table II. Main EFCCs system parameters

| | $R_U$[m] | $Z_U$[m] | $R_L$[m] | $Z_L$[m] |
|---|---|---|---|---|
| Lower array | 3.093 | -0.7737 | 2.975 | -1.072 |
| Equat. array | 3.142 | 0.5251 | 3.142 | -0.5251 |
| Upper array | 2.729 | 1.399 | 3.109 | 0.8190 |

Due to the required accuracy level on the magnetic field (about $1\mu T$), a denser discretization for each DTT

Coil turn has been needed. A dedicated analysis, computing the magnetic field due to PF/TF coils on a set of fixed points and varying the discretization has shown that 3000 segments is the optimal choice (Table III).

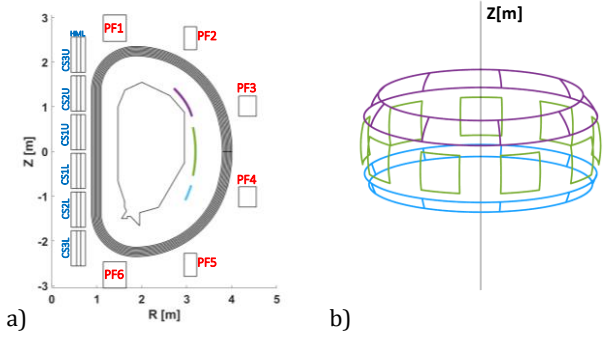

Fig. 1. DTT: a) poloidal Cross section; b) EFCCs system.

Table III. Accuracy level varying the number of sticks

| $N_{seg}$ | 2000 | 3000 | 4000 | 5000 |
|---|---|---|---|---|
| Accuracy [T] | 1.54e-6 | 6.85e-7 | 3.85e-7 | 2.47e-7 |

### 3.2 PF/CS: subset of perturbations

The PF/CS system has been perturbed using two typologies of parameters: i) assembly inaccuracies, modelled as translations respect to $x$, $y$ and $z$ directions, and rotations respect to $\tilde{x}/\!/x$, $\tilde{y}/\!/y$ and $z$ axes, passing through the coil barycenter (Fig. 2); ii) manufacturing errors, modelled as deformations, implemented as an elliptical striation parameter and a coil radius perturbation. Since translations (rotations) respect to $z(\tilde{z})$, radius perturbations, and elliptical deformations do not act on $n=1$ modes, they have no impact on the TMEI and can be disregarded. The $\tilde{x}$ ($\tilde{y}$) axial rotations are implemented using Euler's rotation matrices.

### 3.3 TF: subset of perturbations

TF Assembly inaccuracies, implemented similarly to the PF/CS case, are described in a suitable coordinate system $(\tilde{r},\tilde{\varphi},\tilde{z})$: where $\tilde{r}/\!/r$, $\tilde{z}/\!/z$, and $\tilde{\varphi}=\tilde{z}\times\tilde{r}$ are centered in the TF barycenter (Fig. 3). The non-rigid deformations working on TF D-shape, have been implemented using cubic spline interpolation functions controlled in some points along the D-shape (Fig. 4). All these perturbations have an impact on the TMEI value.

The Spline function is designed to assign the maximum deformation on the controlled nodes. The 1D cubic spline expression, depending on the generic curvilinear abscissa $x$, in the $i$-th interval between two subsequent nodes (with indices $i$ and $i-1$), is:

$$s_3(x) = \frac{(x_i-x)^3 6\left(\frac{f(x_i)-f(x_{i-1})}{h_i^2}\right)+(x_{i-1}-x)^3 6\left(\frac{f(x_i)-f(x_{i-1})}{h_i^2}\right)}{6h_i} \\ + 3\left(\frac{f(x_i)-f(x_{i-1})}{h_i}\right)(x-x_{i-1}) + f(x_{i-1}) - h_i\left(\frac{f(x_i)-f(x_{i-1})}{h_i}\right) \quad (12)$$

where $x_i$ and $x_{i-1}$ are the coordinates of the nodes, $f(x_i)$ and $f(x_{i-1})$ are the respective deformation amplitudes used to control the Spline shape, and $h_i$ is the step.

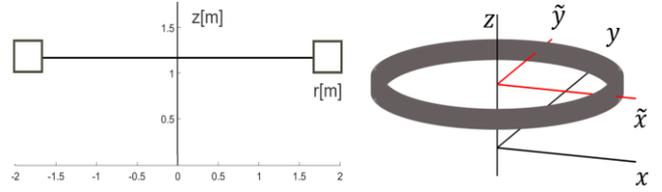

Fig. 2. PF coil section and 3D representations.

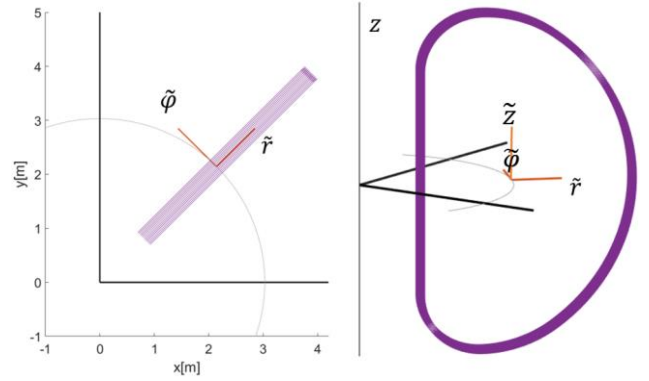

Fig. 3. TF rigid perturbations directions.

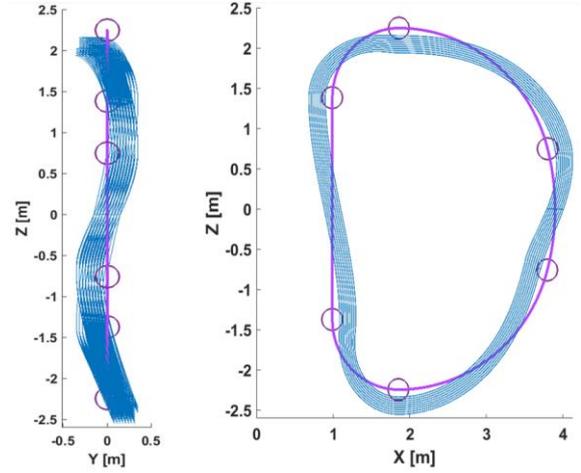

Fig. 4. TF D-shape deformation (the amplitudes have been intentionally exaggerated): The nominal shape (magenta) and 6 control points (circles) are sketched.

## 4. Results

Once the TMEI modes matrices description for all magnetic coils, a fast stochastic analysis can be performed using (9), by randomly varying the set of parameters according to manufacturing and assembling accuracy.

### 4.1 DTT EFs impact: stochastic analysis

An extensive study is planned for the future, aimed at analysing the entire dynamic of the scenarios of interest for DTT. Here the EFs impact has been evaluated, considering CS/PF currents of the Single Null flat-top instant t=36s [13] and TF currents (Table IV).

Table IV. CS/PF and TF currents

| Name | Current [MAt] | Name | Current [MAt] |
|---|---|---|---|
| PF1 | 4.109 | PF4 | -1.815 |
| PF2 | -1.910 | PF5 | -3.211 |
| PF3 | -2.047 | PF6 | 9.331 |
| CS3U-H | -0.07104 | CS3L-H | -1.346 |
| CS3U-M | -0.08358 | CS3L-M | -1.583 |
| CS3U-L | -0.1504 | CS3L-L | -2.850 |
| CS2U-H | 0.3978 | CS2L-H | -0.9327 |
| CS2U-M | 0.4680 | CS2L-M | -1.097 |
| CS2U-L | 0.8424 | CS2L-L | -1.975 |
| CS1U-H | -1.596 | CS1L-H | 0.2946 |
| CS1U-M | -1.878 | CS1L-M | 0.3466 |
| CS1U-L | -3.380 | CS1L-L | 0.6239 |
| TFs (×18) | 3.520 | | |

For the stochastic analysis, $N_{cases} = 1e6$ has been considered and upper bounds for the maximum perturbations have been reported in Table V.

Table V. CS/PF and TF upper perturbations bounds

| | CS [mm] | PF [mm] | TF [mm] |
|---|---|---|---|
| Translations | 2.000 | 4.000 | 4.000 |
| Rotations | 2.000 | 4.000 | 4.000 |
| Deformations | Not present | Not present | 4.000 |

A cumulative distribution function (cdf) has been used to compute 80%, 90% and 95% of the cumulative TMEI values of the probability density function (pdf). Table VI reports the TMEI values for different combinations and Fig. 5 shows the CS+PF+TF pdf.

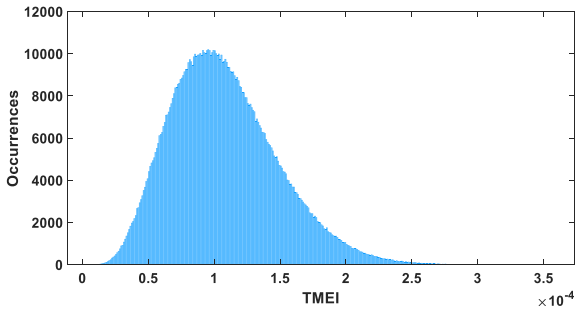

Fig. 5. Pdf of CS+PF+TF contribution to TMEI.

Table VI. TMEI (ppm) values for main percentages of the pdf

| Percentage | CS only | PF only | TF only | CS+PF | CS+PF+TF |
|---|---|---|---|---|---|
| 80% | 30.38 | 39.35 | 132.8 | 46.77 | 140.4 |
| 90% | 35.39 | 45.48 | 154.2 | 53.19 | 162.0 |
| 95% | 39.55 | 50.65 | 172.9 | 58.72 | 180.8 |

*4.2 DTT EFs impact: EFCCs currents limits*

From the previous stochastic analysis, the values of the optimal EFCCs currents have been obtained solving the constrained minimization problem (11) for each case (Fig 6 and Table VII). The analysis shows that 50kAt are able to reduce the TMEI under 50 ppm in 95% of the cases considered, when the CS+PF+TF set of sources is used.

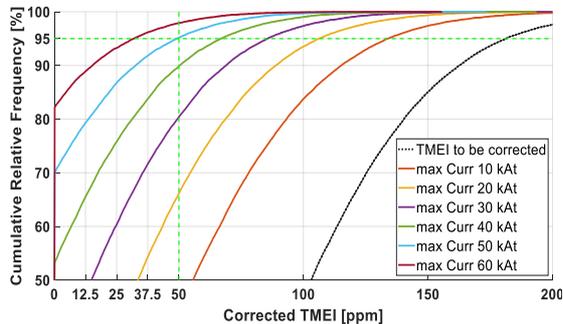

Fig. 6. Cdf curves of TMEI varying currents constraints of EFCCs. The 95% at 50 ppm has been highlighted.

Table VII. TMEI below 50 ppm: percentages of corrected cases varying EFCCs current bounds.

| Currents | CS only | PF only | TF only | CS+PF +TF | TF def. only | TF pert. only |
|---|---|---|---|---|---|---|
| 10 kAt | 100 % | 100 % | 49.78 % | 44.01 % | 80.70 % | 73.21 % |
| 20 kAt | 100 % | 100 % | 70.72 % | 66.12 % | 93.77 % | 90.17 % |
| 30 kAt | 100 % | 100 % | 84.43 % | 80.43 % | 98.35 % | 97.06 % |
| 40 kAt | 100 % | 100 % | 92.41 % | 89.87 % | 99.66 % | 99.28 % |
| 50 kAt | 100 % | 100 % | 96.62 % | 95.18 % | 99.95 % | 99.84 % |
| 60 kAt | 100 % | 100 % | 98.76 % | 97.85 % | 100 % | 99.98 % |

## 5. Conclusions

Error Fields from manufacturing and assembly errors of DTT superconducting coils have been calculated using a stochastic procedure based on a linearized model. The required EFCC currents to correct them below a given threshold TMEI are evaluated by a constrained quadratic programming procedure. EFCC currents of 50 kAt are sufficient to correct the TMEI under 50 ppm with a 95% probability. Further analyses are planned by considering other types of deformations, assessing the values of the maximum expected deformations, and the so-called error field amplifications by the plasma response.

**Acknowledgments**

This paper is partially supported by Italian MUR, PRIN grant #20177BZMAH, in progress among Italian Universities in cooperation with international labs.

This work has been carried out within the framework of the EUROfusion Consortium, funded by the European Union via the Euratom Research and Training Programme (Grant Agreement No 101052200 — EUROfusion). Views and opinions expressed are however those of the author(s) only and do not necessarily reflect those of the European Union or the European Commission. Neither the European Union nor the European Commission can be held responsible for them.

Declaration of Interest Statement

**Declaration of interests**

☐The authors declare that they have no known competing financial interests or personal relationships that could have appeared to influence the work reported in this paper.

☒The authors declare the following financial interests/personal relationships which may be considered as potential competing interests:

Raffaele Martone reports financial support was provided by European Consortium for the Development of Fusion Energy. Andrea Gaetano Chiariello reports financial support was provided by Government of Italy Ministry of Education University and Research.